\documentstyle[12pt]{article}
\begin{document}
\begin{center}{\Large {\bf Evidence of invariance of time scale}}\\ 
{\Large{\bf at critical point in the Ising}}\\
{\Large {\bf meanfield equilibrium equation of state}}\end{center}

\vskip 1 cm

\begin{center}{Muktish Acharyya}\\
{\it Department of Physics}\\
{\it Presidency University, 86/1 College Street}\\
{\it Calcutta-700073, India}\\
{\it muktish.acharyya@gmail.com}\\
{and}\\
{Ajanta Bhowal Acharyya}\\
{\it Department of Physics}\\
{\it Lady Brabourne College}\\
{\it P-1/2 Surahwardy Avenue, Calcutta-700017, India}\\
{\it ajanta.bhowal@gmail.com} \end{center} 
\vskip 1cm

\noindent We solved the equilibrium
meanfield equation of state of Ising ferromagnet (obtained from 
Bragg-Williams theory) by Newton-Raphson method. 
The number of iterations required to
get a convergent solution (within a specified accuracy) 
of equilibrium magnetisation, at any particular temperature, is observed to
diverge in a power law fashion as the temperature approaches the 
critical value. This was identified as the critical slowing down. 
The exponent is also estimated. 
This value of the exponent
is compared with that obtained from analytic solution. 
Besides this, the numerical results are also compared with some 
experimental results exhibiting satisfactory degree of agreement.
It is observed
from this study that the information of the invariance of time scale
at the critical point is present in the meanfield equilibrium equation
of state of Ising ferromagnet.

\vskip 1cm

\noindent {\bf {\it Keywords: Ising model, Meanfield theory, 
Newton-Raphson method, Relaxation time, critical slowing down}}

\noindent {\bf PACS Nos:} 75.40.Mg
\newpage
\noindent {\bf I. Introduction:}

Equilibrium statistical physics took an important place in physics
in last few decades\cite{huang}. Recently, the nonequilibrium 
statistical physics
became an interesting field of modern research. The system far from
equilibrium gives a variety of interesting dynamical phenomena. 
Particularly, the Ising ferromagnet, driven by time dependent
oscillating magnetic field, shows some interesting nonequilibrium 
responses\cite{ma}. 
The attainment
of equilibrium state from an nonequilibrium state gives the transient
behaviour of the system. The relaxation or the transient phenomena of
a cooperatively interacting system is also an important and interesting
subject of study in modern research. Here, mainly the type of relaxation,
the dependence of relaxation time on the temperature, are the main ojective
of research. Generally, the relaxation time is observed to
diverge at the critical point in a power law fashion and this is called
{\it critical slowing down}. This phenomenon was observed 
experimentally in various systems, e.g., 
complex ferromagnetic systems \cite{complex}, ferromagnetic 
iron\cite{ferro}, ferroelectrics \cite{ferroelectrics}, 
spin glass \cite{spinglass} etc. 

In the language of mathematics, this is 
sometimes called the {\it invariance of time scale}, a remarkable feature
of {\it critical phenomena}. On the other hand, the equilibrium equation of
state correctly describes the temperature variation of the order parameter.
In the equilibrium equation of state, generally one cannot expect to get
the information of transient behaviour.

If we want to study the invariance of time scale near the critical point
in any system, we need a time dependent differential equation representing
the transient or nonequilibrium behaviour. Usually, by solving the equation
we have the time dependence of order parameter, which eventually describes
the transient behaviour. From this transient behaviour, the relaxation time
can be calculated and invariance of time scale can be observed at
criticallity. Now the question is, {\it is it possible to study the
nonequilibrium transient behaviour from equilibrium equation of state ?}
We addressed this question in this article and studied the nonequilibrium
behaviour of magnetisation in the equilibrium equation of state of
Ising ferromagnet in meanfield approximation. We have organised the paper
in the following way: in section II we have described the model and
the methodology of solution, section III contains the numerical results,
the paper ends with a summary in section IV.

\vskip 2cm

\noindent {\bf II. Meanfield equation and Numerical Solution:}

We know from the Bragg-Williams theory\cite{huang}
the equilibrium meanfield equation 
of state (in the absence of magnetic field) of Ising ferromagnet becomes
\begin{equation}
m={\rm tanh}({{m} \over {T}}),
\end{equation}

\noindent $m$ is the equilibrium magnetisation and $T$ is the temperature
of the system. Here, the temperature $T$ is measured in such an unit that
the ferromagnetic interaction strength $J$, the
Boltzmann constant $K_B$ and the number of nearest neighbour $q$ are
set equal to unity.

One can solve the equation by Newton-Raphson method \cite{num}to get
the equilibrium value of magnetisation at any fixed temperature. This has
been done already and studied the temperature dependence of magnetisation.
The transition (ferro-Para) temperature $T_c$ can be estimated from this
study and it becomes equal to unity. {\it However, no such 
attempt has been made so far to
study the transient behaviour from this 
equilibrium meanfield equation of state.}

We have solved this equation by Newton-Raphson method
\cite{num} taking $f(m)
= m - {\rm tanh}({{m} \over {T}})$ and by iterating $m_{new} = m_{old}
- {{f(m)} \over {{df} \over {dm}}}$. The criterion of the convergence
of the solution was set as ${{|m_{new} - m_{old}|} \over {m_{new}}} 
\leq \epsilon$. In this present study, we set $\epsilon = 10^{-8}$.
We have set the initial (or guess) value of $m = 1.0$. Our study is mainly
confined in the paramagnetic region ($T > T_c$) and very close to
$T_c = 1.0$.
The number of iterations required to achieve this convergence, is defined
as the relaxation time $T_r$.

\vskip 1cm

\noindent {\bf III. Numerical Results:}

The time variation of magnetisation in kinetic Ising model in meanfield
approximation is governed by the following differential equation\cite{tom}
\begin{equation}
\tau{{dm} \over {dt}} = -m + {\rm tanh}({{m} \over {T}})
\end{equation}

where $m(t)$ is the instantaneous magnetisation, $\tau$ is the microscopic
relaxation time, i.e., the time of single spin flip and $T$ is the
temperature. Just above $T_c$ this equation may be linearised as

\begin{equation}
\tau{{dm} \over {dt}} = -m + {{m} \over {T}}
\end{equation}

which is readily solved to get $m(t) \sim 
{\rm exp}(-{{(T-1)t} \over {T\tau}})$ and the relaxation is exponential
with the relaxation time $T_r \sim (T-1)^{-1}$. Here, the analytical 
value of $z$ is equal to 1.0 and $T_c=1.0$ 

We have calculated this relaxation time
$T_r$ for different temperatures $T$. This is calculated from the
number of iterations required to get the convergent
(within accuracy $\epsilon = 10^{-8}$)
 solution of
eqn(1) by Newton-Raphson method.
Let us call this is the relaxation time $T_r$ calculated from convergent
solution.
Figure-1(a) shows the temperature
variations of relaxation times $T_r$
(obtained from convergence). It is clear that The relaxation time
$T_r$ shows the tendency of divergence as one approaches $T_c=1.0$. This
is a clear indication of {\it critical slowing down} or the {\it invariance
of time scale}. Amazingly, this is being observed here from the 
equilibrium equation
of state where the information of time variation is absent!
The relaxation time $T_r$ can also be measured in a different way. We have
recorded the value of $m$ obtained at each iteration in solving the
meanfield equation by Newton-Raphson method. 
If this is plotted against the number of iterations
this shows an exponential variation. Figure-1(b) shows the semilog plot of
magnetisation ($m(i)$) after i-th iteration versus the 
number of iteration $i$.
The straight line in semilog plot reveals the exponential nature of 
relaxation. Assuming $m(i) \sim {\rm exp}(-i/T_r)$, the relaxation time
$T_r$ is calculated from the slope of semilog plot of $m(i)$ versus $i$.
Let us call this is the relaxation time $T_r$ calculated from the 
exponential relaxation.
Here also the relaxation time $T_r$ increases as the temperature $T$ comes
closer to $T_c = 1.0$ and shows also a tendency of divergence near $T_c=1.0$.
Here also we observed the similar kind of 
{\it critical slowing down} or {\it invariance of time scale}. 
This is shown in Figure-1(c).

Our next step is to estimate the exponent assuming the power law
divergence of relaxation time. 
First we assume the scaling law
$T_r \sim (T-T_c)^{-z}$. For the right choice of $z$, the plot of
$T_r^{-1/z}$ versus $T$ will be a straight line. The straight line
will intersect the $T$-axis at $T=T_c$. We have chosen a guess value
of $z$ and fitted $T_R^{-1/z}$ versus $T$ with a straight line
by least square fitting method. We have calculated the error 
$e_r$ in fitting. Now we varied $z$ systematically in a range 
of values and
calculated the error (in least square straight line fitting) $e_r$.
The error $e_r$ is plotted against $z$ and shown in fig-2(a). We have
chosen the value of $z$ which minimises the error $e_r$. This becomes
$z=0.989$ and with this value the straight line fit is also shown in 
fig-2(b). This straight line cuts the $T$ axis at $T=T_c=1.0000$. 

The same technique was employed to estimate $z$ and $T_c$, in the case
where we have calculated the relaxation time $T_r$ by exponential 
relaxation method. This is shown in fig-3. Here, the exponent $z$=0.990
and $T_c=1.0000$.

Both the results of $z$ and $T_c$ estimated here (even by both methods
of calculating the relaxation time $T_r$) agree well with that obtained
analytically, i.e. $z=1.0$ and $T_c=1.0$. 

It may be mentioned here that the experiments\cite{complex}
 on the spin dynamics
near the critical point was performed in quasi two dimensional ferromagnet
$(CH_3 NH_3)_2 Cu Cl_4$. The similar power law variation of relaxation time
was observed and the exponent
estimated is $1.05 \pm 0.03$ which agrees well with that obtained in our
present meanfield calculation.

One may argue that this may be some artefact of Newton-Raphson method.
Keeping this in mind we have applied simple iterative method and calculated
the relaxation time from the convergence criterion. This also shows 
similar result but the exponent estimated here is different $z=0.857$, 
however the same power law divergence of relaxation time was observed.
These resuls are shown in Figure-4.
\vskip 1cm

\noindent {\bf IV. Summary:}

In this paper, we studied mainly the transient behaviour from
time independent equilibrium equation of state, particularly,
the equation of state of Ising ferromagnet in meanfield approximation.
The equation of state, gives the equilibrium magnetisation at
any particular temerature and can be obtained by solving the
transcendental equation by Newton-Raphson method. Generally,
the convergent solution gives the equilibrium result. However,
the transient behaviour may be obtained from this method.
Here, we have calculated the relaxation time, defined as the
number of iterations required to get the convergent solution,
and studied it as a function of temperature. 
This relaxation time was observed to diverge at the critical point
in a power law fashion and the exponent is also estimated. The estimated
value of the exponent agrees well with that obtained analytically. The
critical slowing down and the power law divergence of the relaxation time
are also observed experimentally. Our meanfield estimate of the 
exponent agrees well with that obtained experimentally\cite{complex}.

In conclusion, let us say that the {\it transient} behaviour
may be obtained from the {\it equilibrium equation of state}. The time
independent equilibrium equation of state contains the information
about the transient behaviour. 

One possible explanation of this fact may be the method of getting the
relaxation behaviour. Both Newton-Raphson and simple iterative methods
are Markovian type and the differential equation of getting the time
dependent magnetisation is of first order.

\vskip 1cm

\noindent {\bf V. Acknowledgements:}

We would like to thank Subinoy Dasgupta for helpful discussions.
We also thank to Abhik Basu for helping us to collect Reference- \cite{tom}.

\newpage
\setlength{\unitlength}{0.240900pt}
\ifx\plotpoint\undefined\newsavebox{\plotpoint}\fi
\sbox{\plotpoint}{\rule[-0.200pt]{0.400pt}{0.400pt}}%


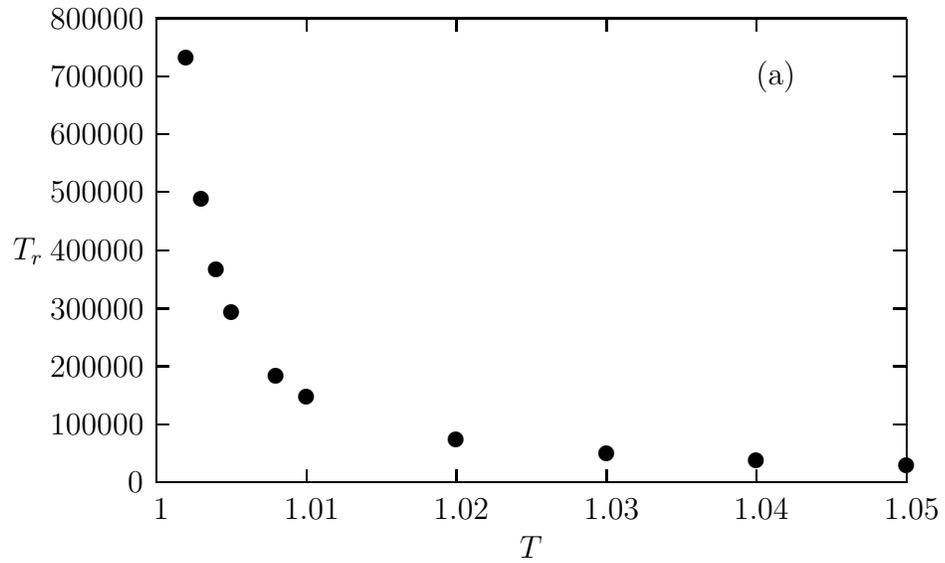
\noindent {\bf Fig.-1(a).} The variation of relaxation
time $T_r$ against the temperature $T$. The relaxation
time is calculated from the convergence time of solving
$m$ (from eqn(1)) by Newton-Raphson method. 

\newpage

\setlength{\unitlength}{0.240900pt}
\ifx\plotpoint\undefined\newsavebox{\plotpoint}\fi
\sbox{\plotpoint}{\rule[-0.200pt]{0.400pt}{0.400pt}}%
%

\vskip 1cm

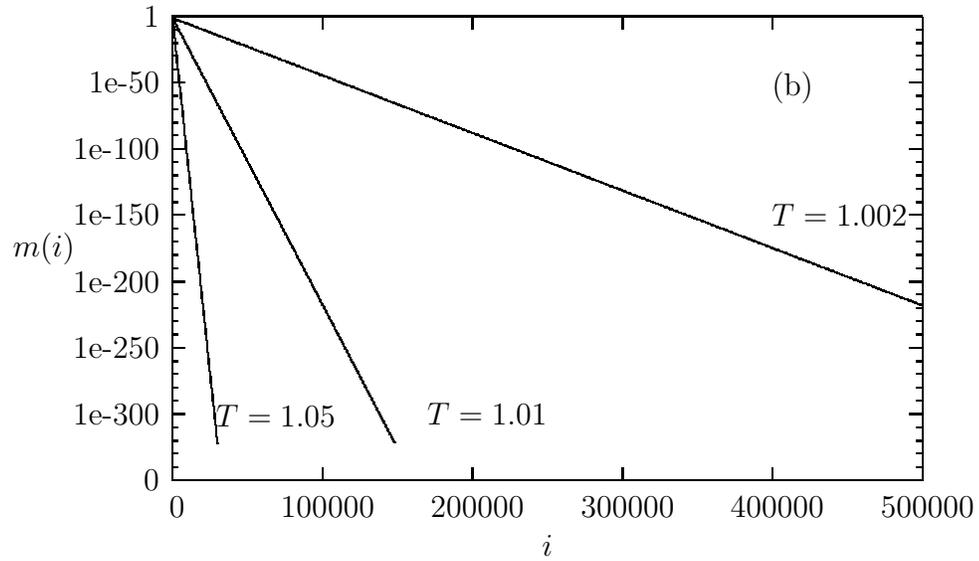
\noindent {\bf Fig.-1(b).} The evidence of exponential relaxation. 
The semilog plot of magnetisation $m(i)$ (value of $m$ at i-th
iteration in solving eqn(1)) 
versus time $i$ (i-th iteration) at different 
temperatures.

\newpage
%

\vskip 1cm

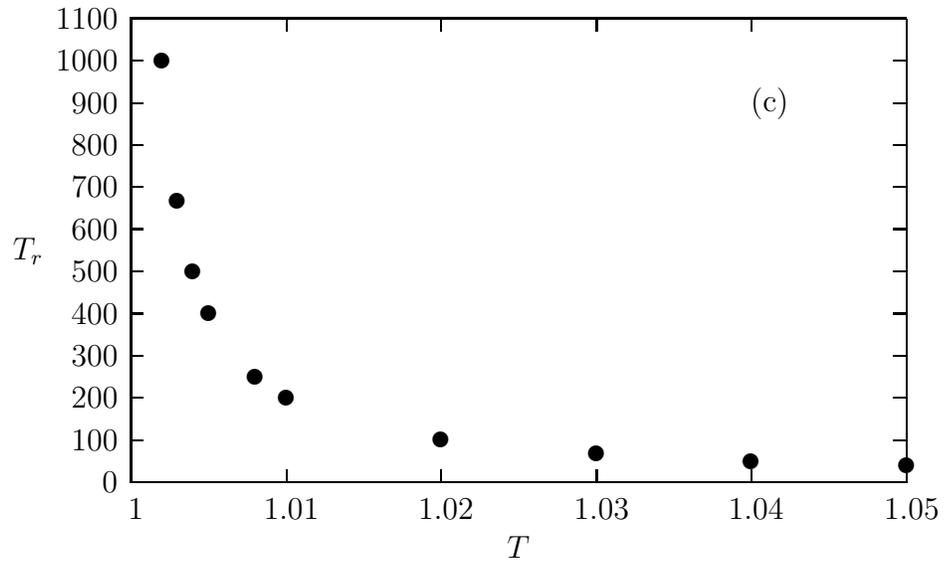
\noindent {\bf Fig.-1(c).} The variations of relaxation time ($T_r$)
with temperature ($T$).Here, the relaxation time $T_r$ is calculated
from the exponential fitting of the solution $m(t)$ of eqn(1).

\newpage

\setlength{\unitlength}{0.240900pt}
\ifx\plotpoint\undefined\newsavebox{\plotpoint}\fi
\sbox{\plotpoint}{\rule[-0.200pt]{0.400pt}{0.400pt}}%
%

\vskip 1cm

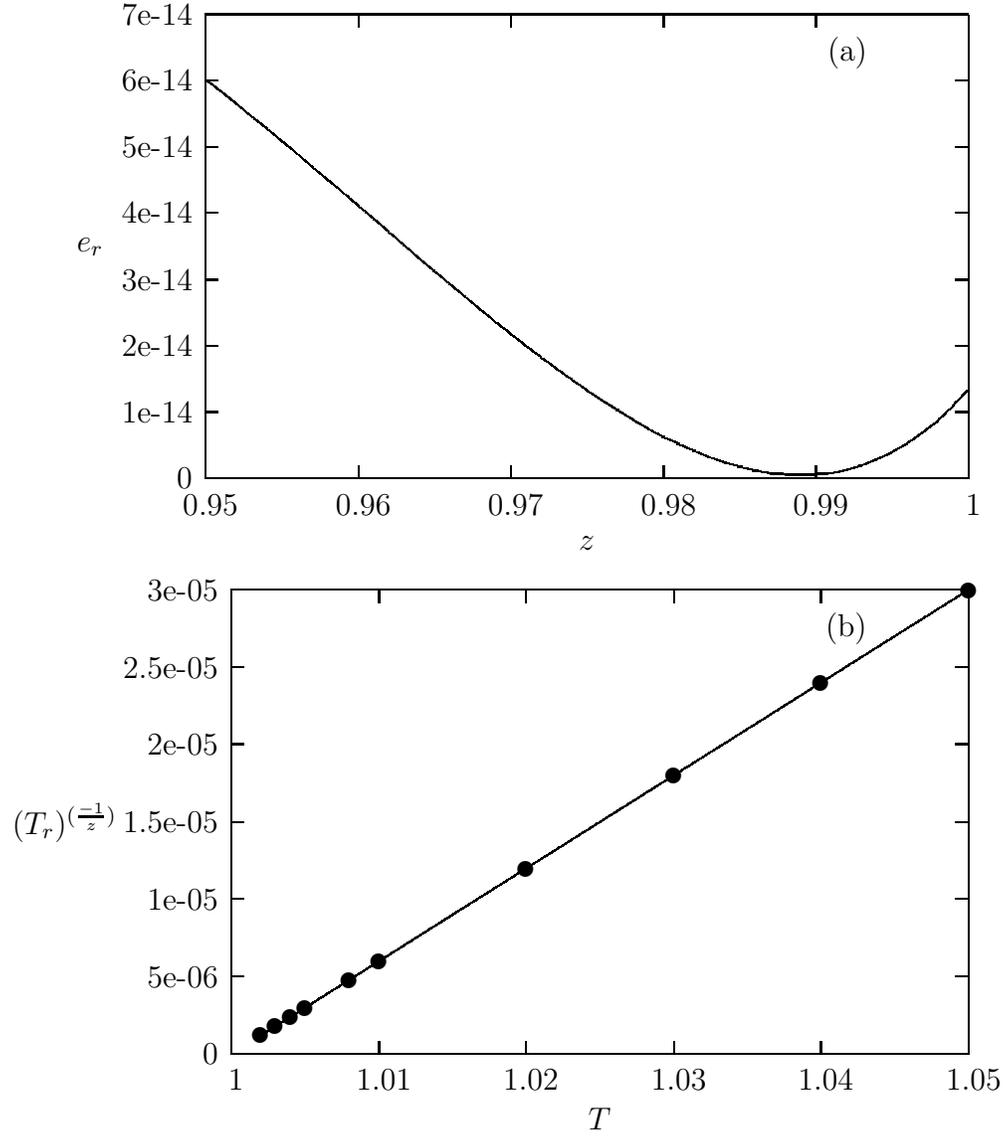
\noindent {\bf Fig.-2.} (a) The error is plotted against the various
values of exponent $z$. The error becomes minimum for $z=0.989$. (b)
the plot of $(T_r)^{({{-1} \over {z}})}$ ($z=0.989$) against the 
temperature $T$. Continuous line is the linear best fit. Here the
estimated $T_c = 1.0000$. The $T_r$ obtained here from convergence criterion.

\newpage

\setlength{\unitlength}{0.240900pt}
\ifx\plotpoint\undefined\newsavebox{\plotpoint}\fi
\sbox{\plotpoint}{\rule[-0.200pt]{0.400pt}{0.400pt}}%
%

\vskip 1cm
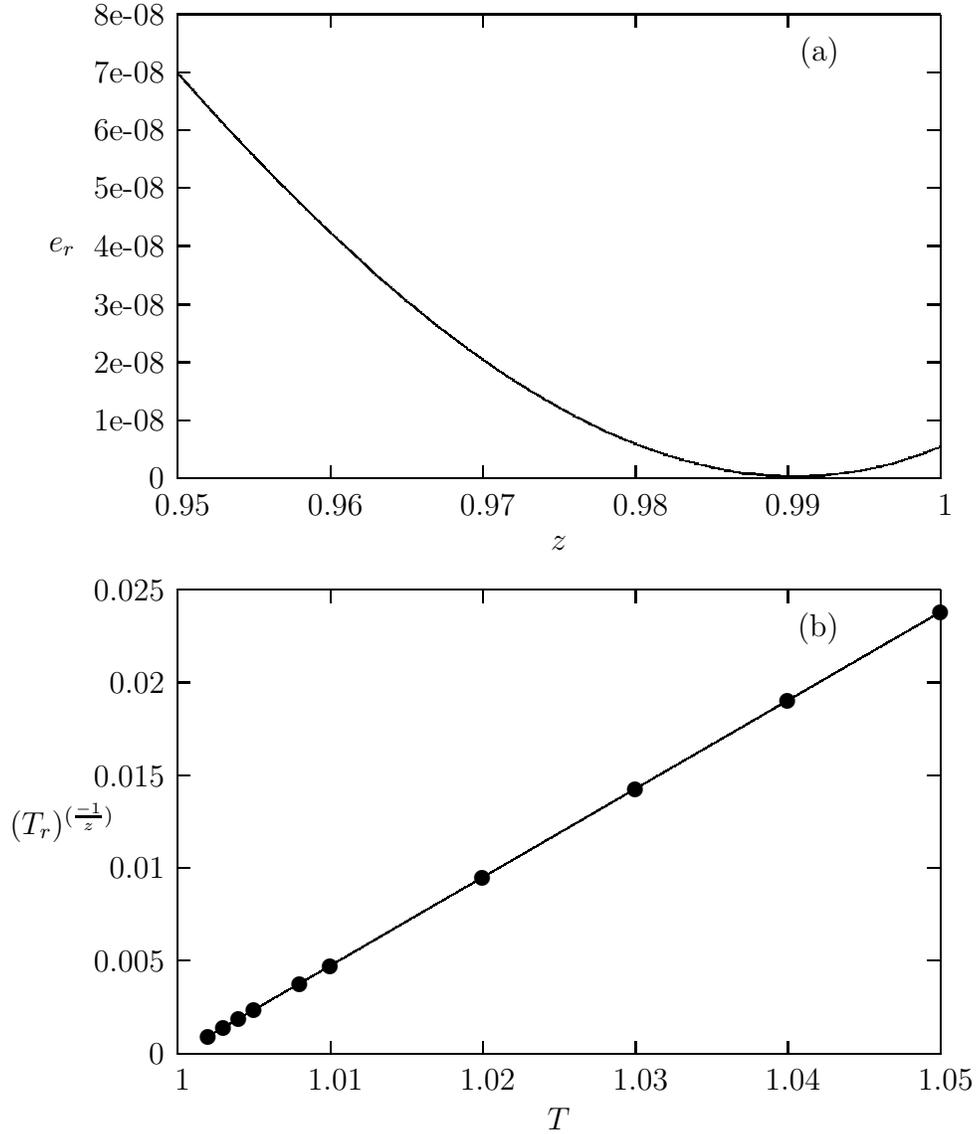
\noindent {\bf Fig.-3.} (a) The error is plotted against the various
values of exponent $z$. The error becomes minimum for $z=0.990$. (b)
the plot of $(T_r)^{({{-1} \over {z}})}$ ($z=0.990$) against the 
temperature $T$. Continuous line is the linear best fit. Here the
estimated $T_c = 1.0000$. The $T_r$ obtained here from exponential 
relaxation.

\newpage

\setlength{\unitlength}{0.240900pt}
\ifx\plotpoint\undefined\newsavebox{\plotpoint}\fi
\sbox{\plotpoint}{\rule[-0.200pt]{0.400pt}{0.400pt}}%
%

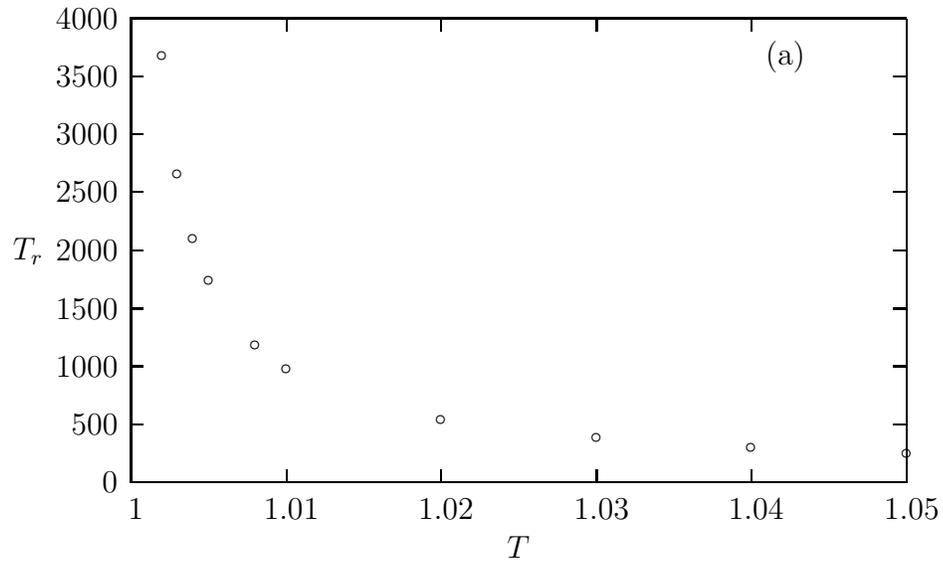
\noindent {\bf Fig.4a.} The relaxation time ($T_r$) plotted against
temperature ($T$). Here, $T_r$ is calculated from the number of
iteration required for getting convergent solution on $m$ by iterating
the equation $m = {\rm tanh}(m/T)$.

\newpage

\setlength{\unitlength}{0.240900pt}
\ifx\plotpoint\undefined\newsavebox{\plotpoint}\fi
\sbox{\plotpoint}{\rule[-0.200pt]{0.400pt}{0.400pt}}%
%

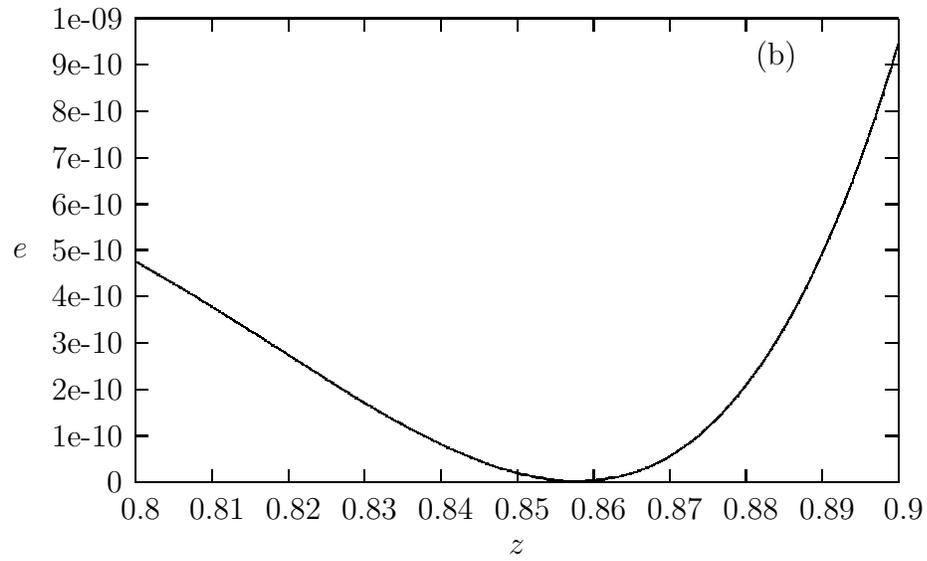
\noindent {\bf Fig.4b.} The error ($e$) of linear fitting is plotted 
against test values of exponent ($z$). The error is minimum for
$z=0.857$.
\newpage

\setlength{\unitlength}{0.240900pt}
\ifx\plotpoint\undefined\newsavebox{\plotpoint}\fi
\sbox{\plotpoint}{\rule[-0.200pt]{0.400pt}{0.400pt}}%
%

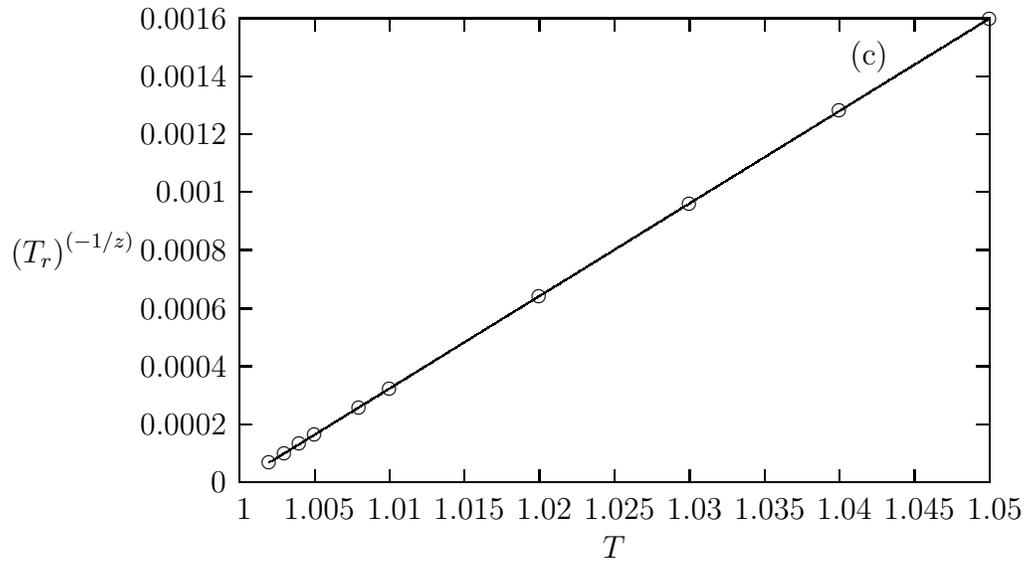
\noindent {\bf Fig.4c.} The plot of ${(T_r)}^{(-1/z)}$ versus temperature
($T$). Continuous line is linear best fit with $z=0.857$.

\newpage

\noindent {\bf References}
\begin{enumerate}
\bibitem{huang} K. Huang, Statistical mechanics, Wiley, 
John and Sons, 1990
\bibitem{ma} B. K. Chakrabarti and M. Acharyya, Rev. Mod. Phys., 71 (1999)
847 and the references therein.
\bibitem{complex} Y. Okuda et al., J. Phys. Soc. Jpn., 47 (1979) 773
\bibitem{ferro} M. J. Dunlavy and D. Venus, Phys. Rev. B, 69 (2004) 094411
\bibitem{ferroelectrics} E. Kanda et al., J. Phys. C: Solid State Physics,
15 (1982) 3401
\bibitem{spinglass} N. S. Sullivan and D. Esteve, Physica B+C 107 (1981) 189
\bibitem{tom} T. Tome and M. J. de Oliveira, Phys. Rev. A 41 (1990) 4251;
See also, R. Kubo, J. Phys. Soc. Jpn. 24 (1968) 51,
R. J. Glauber, J. Math. Phys. 4 (1963) 294.
\bibitem{num} B. Scarborough, Numerical Mathematical Analysis, 
Oxford and IBH, (1930).
\end{enumerate}
\end{document}